\documentclass[sigconf]{acmart}

\AtBeginDocument{%
  }

\usepackage{array} % for m{} and column modifiers

\usepackage{booktabs}  % for \toprule, \midrule, \cmidrule, \bottomrule

\usepackage{subcaption}
\usepackage{siunitx}
\copyrightyear{2026}
\acmYear{2026}
\setcopyright{cc}
\setcctype{by}
\acmConference[LAK 2026]{LAK26: 16th International Learning Analytics and Knowledge Conference}{April 27-May 01, 2026}{Bergen, Norway}
\acmBooktitle{LAK26: 16th International Learning Analytics and Knowledge Conference (LAK 2026), April 27-May 01, 2026, Bergen, Norway}
\acmPrice{}
\acmDOI{10.1145/3785022.3785105}
\acmISBN{979-8-4007-2066-6/2026/04}

\begin{document}
%%\linenumbers

%%
%% The "title" command has an optional parameter,
%% allowing the author to define a "short title" to be used in page headers.
\title{Structuring versus Problematizing: How LLM-based Agents Scaffold Learning in Diagnostic Reasoning}

%%\author{Anonymous Author(s)}

\author{Fatma Betül Güreş}
\orcid{0000-0002-8664-1022}
\affiliation{%
  \institution{ETH Zürich, EPFL}
  \city{Lausanne}
  \country{Switzerland}
}
\email{fatma-betul.gures@epfl.ch}

\author{Tanya Nazaretsky}
\orcid{0000-0003-1343-0627}
\affiliation{%
  \institution{EPFL}
  \city{Lausanne}
  \country{Switzerland}
}
\email{tanya.nazaretsky@epfl.ch}

\author{Seyed Parsa Neshaei }
\orcid{0000-0002-4794-395X}
\affiliation{%
  \institution{EPFL}
  \city{Lausanne}
  \country{Switzerland}
}
\email{seyed.neshaei@epfl.ch}

\author{Tanja Käser}
\orcid{0000-0003-0672-0415}
\affiliation{%
  \institution{EPFL}
  \city{Lausanne}
  \country{Switzerland}
}
\email{tanja.kaeser@epfl.ch}

\renewcommand{\shortauthors}{Güreş et al.}

%%
%% The abstract is a short summary of the work to be presented in the
%% article.
\begin{abstract}
Supporting students in developing diagnostic reasoning is a key challenge across educational domains. Novices often face cognitive biases such as premature closure and over-reliance on heuristics, and they struggle to transfer diagnostic strategies to new cases. Scenario-based learning (SBL) enhanced by Learning Analytics (LA) and large language models (LLM) offers a promising approach by combining realistic case experiences with personalized scaffolding. Yet, how different scaffolding approaches shape reasoning processes remains insufficiently explored. This study introduces PharmaSim Switch, an SBL environment for pharmacy technician training, extended with an LA- and LLM-powered pharmacist agent that implements pedagogical conversations rooted in two theory-driven scaffolding approaches: \emph{structuring} and \emph{problematizing}, as well as a student learning trajectory. In a between-groups experiment, 63 vocational students completed a learning scenario, a near-transfer scenario, and a far-transfer scenario under one of the two scaffolding conditions. Results indicate that both scaffolding approaches were effective in supporting the use of diagnostic strategies. Performance outcomes were primarily influenced by scenario complexity rather than students' prior knowledge or the scaffolding approach used. The structuring approach was associated with more accurate Active and Interactive participation, whereas problematizing elicited more Constructive engagement. These findings underscore the value of combining scaffolding approaches when designing LA- and LLM-based systems to effectively foster diagnostic reasoning. 
\end{abstract}



%% A "teaser" image appears between the author and affiliation
%% information and the body of the document, and typically spans the
%% page.
\maketitle

\section{INTRODUCTION}
Supporting learners in acquiring complex reasoning skills remains a central challenge across professional and academic domains \cite{van2018ten,Lazonder2016}. Among these, diagnostic reasoning is particularly critical: it enables practitioners to evaluate evidence, generate and test hypotheses, and make informed decisions under uncertainty. Such skills are required in a range of domains, such as healthcare, engineering, business, and education \cite{bowen2006, delavari2024teaching,gures2025seq}. Yet, research shows that novices often rely on surface-level heuristics, make premature decisions, or struggle to transfer strategies across contexts \cite{barnett2002when,croskerry2003}. Scenario-Based Learning (SBL) environments have emerged as a promising instructional approach, immersing learners in realistic and open-ended tasks that mirror the ambiguity of real practice \cite{Clark2012}.  However, these environments are not without challenges. If left unguided, learners may experience cognitive overload, fail to explore alternatives, or settle too quickly on an initial idea \cite{han2022faq}. This mismatch between the potential of SBL and the actual learning outcomes underscores the importance of integrating scaffolding that can guide and sustain productive engagement \cite{Reiser2001,Ertmer2019,Adler2023}.  

Recent advances in large language models (LLMs) have created new opportunities to embed conversational agents into SBL environments \cite{radmehr2025pharmasimtext,fassler2023problem}. These agents can act as \textit{mentors}, by posing questions, offering feedback, and tailoring prompts to learners’ progress, providing support that is both adaptive and immediate \cite{Neshaei2025reflective,Adler2023,Reiser2001}. Early research suggests that such agents can scaffold reflection \cite{Neshaei2025reflective} and encourage learners to adopt more systematic strategies \cite{sinha2021}. Yet, the specific ways in which different types of scaffolding are assisted through LLMs, and their effects on learners’ reasoning processes, remain open questions.  
%To date, much of the work applying GPT-based models in medical and healthcare education has focused on simulating patients or generating scenarios and assessment materials. In contrast, our work examines LLMs as mentor-style agents that engage learners in pedagogical conversations, where students can ask clarifying questions, articulate their reasoning, and receive formative, theory-driven feedback. Such interactive mentoring is difficult to realize with traditional scaffolding approaches and remains relatively underexplored in prior LLM-based educational research.
Two scaffolding approaches, both strongly grounded in learning sciences and the theories of student engagement in learning analytics \cite{Quintana2004AInquiry,Chi2009}, are particularly relevant in this context: \textit{structuring} and \textit{problematizing} \cite{Reiser2004ScaffoldingWork,Wood1976,Collins1989}. Structuring scaffolding organizes learners’ work by making expert processes explicit, breaking down tasks, and focusing attention on key steps. This can reduce ambiguity and help learners progress effectively. Problematizing scaffolding, in contrast, deliberately introduces uncertainty and challenges learners to articulate, justify, and reflect on their reasoning. While structuring is expected to improve task accuracy and efficiency, problematizing is meant to foster deeper reasoning through productive struggle. %, but little is known about their comparative impact when implemented in authentic reasoning tasks through LLM-powered agents.  
Despite increasing interest in the role of both types of scaffolding in AI-supported learning environments, exploring their design in human tutoring and computer-based instruction \cite{Reiser2001,Quintana2004AInquiry,Roll2018}, empirical evidence on their \textit{comparative} impact on shaping learners’ reasoning strategies, interaction behaviors, and transfer of knowledge remains scarce.
% Prior work has emphasized the theoretical benefits of structuring and problematizing scaffolds \cite{} and explored their design in human tutoring and computer-based instruction \cite{}. However,
%We still lack systematic studies that evaluate how these scaffolding approaches differ in shaping learners’ reasoning strategies, interaction behaviors, and transfer of knowledge when implemented in complex SBL environments through LLM-driven agents. Understanding these dynamics is crucial for advancing both the design of adaptive scaffolding and the broader integration of AI into professional education.

To address these gaps, this paper introduces PharmaSim Switch, a scenario-based simulation designed to support pharmacy apprentices in developing diagnostic reasoning skills. In our study, 63 vocational school students engaged with the system while receiving either structuring or problematizing scaffolding from an LA- and LLM-powered pharmacist character. We examined how these scaffolding approaches influenced diagnostic outcomes, interaction behaviors, and the transfer of reasoning across scenarios of varying complexity. %We also explored whether prior conceptual and strategy knowledge moderated these effects.
% Parsa: check here to make sure it's true
%Our study utilizes learning analytics to design rubric-based evaluations grounded in prior research, aiming to evaluate the pedagogical effectiveness of AI-supported scaffolding. 
We analyze how LA- and LLM-based agents influence learners’ strategies, engagement, and reasoning processes, and investigate their impact on learning in real-world classrooms. Based on our design of PharmaSim Switch, we addressed the following research questions:  
\textbf{RQ1:} To what extent can LLM-based agents for diagnostic strategies provide different types of scaffolding to learners in simulated scenarios?
% \textbf{RQ1:} To what extent do Structuring-heavy and Problematizing-heavy agents differ in the types of scaffolding they provide and the diagnostic strategies they emphasize?  
\textbf{RQ2:} How do Structuring- and Problematizing-heavy agents affect students’ learning outcomes in acquiring diagnostic strategies across learning phase and transfer phases, such as near and far transfer contexts?  
\textbf{RQ3:} How do students’ behavioral patterns differ between scaffolding conditions?

%Our findings indicate that LLM-based agents embedded in pharmacy simulations successfully achieved the intended scaffolding approaches, with both structuring and problematizing behaviors closely aligning with their respective pedagogical designs. We did not find significant differences in overall learning outcomes between the two groups. Our analysis of experimental data showed that problematizing scaffolding led to more constructive engagement, with students articulating and justifying their reasoning with a higher frequency but also with a higher rate of error. However, structuring scaffolding prompted more accurate engagement and higher interaction fluency.

Our work contributes to LA literature by (1) empirically showing how LA- and LLM-based agents can support monitoring student learning trajectory while interacting with different pedagogical scaffolding approaches, namely structuring and problematizing, in scenario-based learning environments, (2) analyzing how these scaffolds shape learners’ diagnostic reasoning performance and transfer across a set of scenarios of various complexities, and (3) using fine-grained student-pharmacist discourse data to classify engagement patterns based on the ICAP \cite{Chi2009} and \citet{Reiser2004ScaffoldingWork} frameworks and measuring the correctness and pedagogical alignment of agent responses. By analyzing how LA and LLM-driven scaffolding affects learner behavioral processes in real-world vocational classroom settings, we contribute to the development of analytics that explain and evaluate pedagogically aligned LA-based tools and simulations for scenario-based learning.

\section{BACKGROUND}

\textbf{Diagnostic Strategies.}
Diagnostic reasoning involves two key processes: data collection and data interpretation, which together support systematic, evidence-based decision-making \cite{elstein1978,gures2025seq}. Structured data collection reduces premature closure \cite{kassirer1991} and improves diagnostic accuracy by ensuring all plausible causes are considered \cite{graber2005diagnostic}. In data collection, a \textit{structured checklist} is a key strategy for comprehensive symptom assessment, fostering systematic questioning and reducing reliance on intuition \cite{gawande_checklist}. A typical implementation is the LINDAFF framework, widely used in Swiss vocational schools for pharmacy technicians, which guides evaluation of symptom characteristics and related factors \footnote{\url{https://www.pharmawiki.ch/wiki/index.php?wiki=LINDAAFF}}. Beyond symptoms, diagnostic reasoning also requires attention to external influences, such as family and caregivers, which are addressed through the \textit{interpersonal relationships strategy}  \cite{Campbell2014}. Next, the \textit{data interpretation strategy} involves generating, evaluating, and prioritizing diagnostic hypotheses \cite{Graber2012}. This process requires systematically compiling potential causes based on symptoms and patient information, then assessing their likelihoods by weighing available evidence. To conclude, diagnostic reasoning is an iterative process in which data collection and interpretation cycles continuously refine understanding \cite{croskerry2003}. Initial interpretations often reveal gaps that prompt further inquiry, with the final decision reached by ruling out less probable explanations and converging on the most likely diagnosis. This complexity makes teaching and learning demanding, underscoring the need for personalized scaffolding.

\noindent\textbf{Structuring and Problematizing Scaffolding.}
\label{ssec:structuring-and-problematizing-scaffolding-rel-work}
Building on classic accounts of scaffolding as a \emph{reduction in degrees of freedom} \cite{Wood1976}, \citet{Reiser2004ScaffoldingWork} conceptualizes scaffolding as two complementary functions: \emph{structuring} student work to make complex tasks tractable, and \emph{problematizing} student work to press learners to engage with core disciplinary ideas rather than skimming the surface. %, structuring support organises activities so that novices can proceed productively without being overwhelmed; conversely, problematizing deliberately makes key aspects of the work \emph{more} salient (and sometimes more demanding) to promote deeper sense-making \cite{Reiser2004ScaffoldingWork, Engle2002}.
Structuring mechanisms typically operate through three recurring mechanisms: (i) \emph{decomposing complex tasks}: breaking the activity into sequenced, manageable subgoals and making plans explicit; (ii) \emph{focusing students' efforts}: highlighting what to attend to and when; and 
(iii) \emph{monitoring and regulating progress}: making progress visible and cueing next steps \cite{Reiser2004ScaffoldingWork, Quintana2004AInquiry}. 
%In the context of diagnostic reasoning, these mechanisms align with prompts that scaffold checklists, prioritisation, and stepwise data collection \cite{Graber_checklists, Gawande_checklist, Mamede2014can, bowen2006}.
While problematizing mechanisms scaffold learning by pressing on the conceptual work through: (i) \emph{eliciting articulation}: making reasoning explicit; (ii) \emph{eliciting decisions}: committing to judgments among constrained options; and (iii) \emph{surfacing gaps and disagreements}: making inconsistencies and unresolved claims visible for resolution \cite{Reiser2004ScaffoldingWork}. Such moves increase attention to epistemic features of practice (e.g., evidence–claim relations or justifications), increasing productive struggle \cite{Quintana2004AInquiry}.
In this study, we adopt Reiser's framework as the theoretical basis for our LLM mentors. The \emph{Structuring-heavy} agent predominantly operationalizes decomposing, focusing, and monitoring, while the \emph{Problematizing-heavy} agent predominantly operationalizes eliciting articulation and decisions, as well as surfacing gaps. 

% The concrete mapping from mechanisms to prompt families and strategy-specific cues is detailed in Section~\ref{subsec:measurement-analysis} and Table~\ref{tab:scaffolding-overview}.

\label{sec:pharmasim_switch}

\subsection{PharmaSim Switch - Modules} 
\begin{figure*}[t!]
    \centering
    \includegraphics[width=\linewidth]{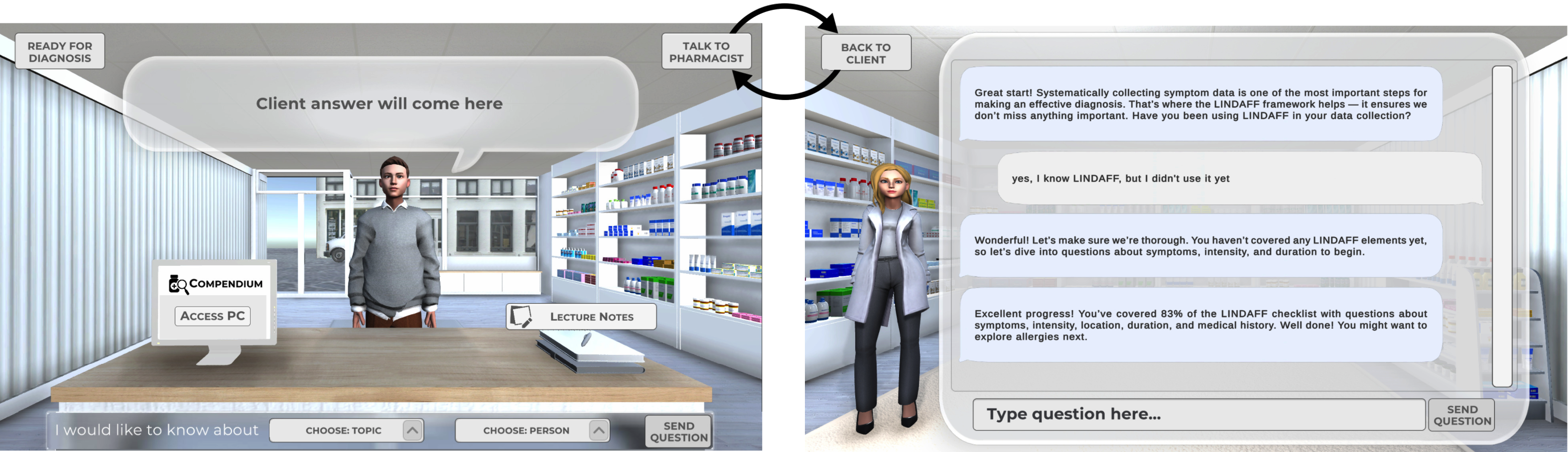}
    \caption{PharmaSim Switch Client Inquiry and Research Module (Left) and Pedagogical Module  (Right)}
    \label{fig:scaffolding}
\end{figure*}

\noindent\textbf{Chatbots and LLMs in Education.}
Conversational agents (e.g., Chatbots) have long been explored as a means of supporting learning, from early rule-based tutors to contemporary dialogue systems \cite{Winkler2020Systematic}. With the advent of large language models (LLMs), chatbots have become capable of generating adaptive, open-ended dialogue with minimal effort \cite{Kasneci2023ChatGPT,Zhang2023ChatGPT}. Recent studies highlight their potential to scaffold reflection, provide formative feedback, and adapt to learners' needs in real time \cite{Chen2024, Neshaei2025reflective}. At the same time, research highlights challenges such as partial coverage of learning goals, low alignment to learners' knowledge states, low reliability, susceptibility to bias, and a lack of transparency in how scaffolding is supported \cite{Luckin2023}. Within learning analytics, chatbots are increasingly integrated into dashboards and feedback systems to support sense-making and strategy use \cite{Noroozi2024,khosravi2023generative,10.1145/3706468.3706545}. However, empirical studies that systematically compare distinct scaffolding approaches supported by LA- and LLM-driven agents remain rare. Our work addresses this gap by investigating how LA- and LLM-powered mentors embody structuring versus problematizing scaffolds in scenario-based learning.

\section{PharmaSim Switch - Pharmacy Assistant Training with Structuring and Problematizing Scaffolding}

PharmaSim is a scenario-based learning (SBL) environment designed to train pharmacy apprentices in diagnostic reasoning by engaging them with a variety of client scenarios \cite{radmehr2025pharmasimtext,gures2025seq}. PharmaSim comprises two main modules: the Client Inquiry and Research module and the Diagnostic Decision module. To explore how different scaffolding strategies shape learning, we enhanced PharmaSim with a Pedagogical Module, an Interactive pharmacist agent that promotes the development of students’ diagnostic reasoning, resulting in the extended environment we call PharmaSim Switch. Below, we provide a detailed explanation of each module and the architecture of the novel Pedagogical module.

The \textit{Client Inquiry and Research Module} aims to replicate the client consultation process (Fig. \ref{fig:scaffolding}, left) and allow data collection. In this module, students interact with the system through drop-down menus to select the individual (e.g., the main client or relatives) and the subject of inquiry (e.g., symptoms, age, allergies), enabling them to pose targeted questions and gather information from the client and relevant relatives. Additionally, students can access two auxiliary knowledge sources: a digital compendium containing the medicine's references and lecture notes, which provide relevant theoretical knowledge. The novel \textit{Pedagogical Module} (Fig. \ref{fig:scaffolding}, right) is implemented as a natural language conversation with a pharmacist, who provides the student with personalized guidance through identifying possible causes. The pedagogical module is implemented in two pedagogical variants: structuring and problematizing scaffolding. Movement between modules could be freely initiated by the student or triggered automatically by the system. This iterative interaction design allows students to consult with the pharmacist at any stage of the diagnostic process, revisiting the client inquiry or resources as needed, and interpret the data. In this way, the system supports a dynamic learning flow, where students alternate between collecting information, consulting knowledge resources, and refining their diagnostic reasoning with pedagogical guidance. Finally, when students decide they have gathered sufficient data, they can transition to the \textit{Diagnostic Module}, where they complete the data interpretation by identifying and justifying potential causes and assessing their likelihood, mirroring the critical thinking and analytical processes used in real-world clinical practice. After providing their answers, students receive the correct solution in the form of a table listing the possible causes of the case, along with their supporting factors and assessed likelihoods.

\subsection{PharmaSim Switch - Pedagogical Module Architecture} 
\label{ssec:chatbot-algorithm}
\begin{figure*}[t!]
\centering
  \includegraphics[width=\linewidth]{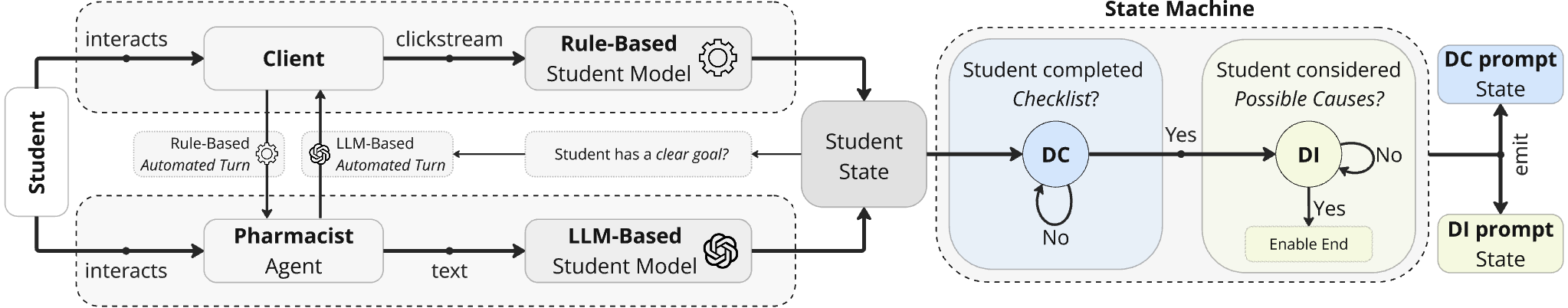}
  \caption{The architecture of models behind PharmaSim Switch. The student interacts with the client and pharmacist characters. The client character is based on a rule-based model connected to a pre-defined knowledge base, while the pharmacist character is an LLM-based agent. Specific student model agents for data collection (DC) and data interpretation (DI) manage moving between the states in the pharmacist agent and the different clients in PharmaSim Switch, respectively.}
  \Description{Methodology}
  \label{fig:arch-models}
\end{figure*}

\begin{figure*}[t]
\centering
  \includegraphics[width=\linewidth]{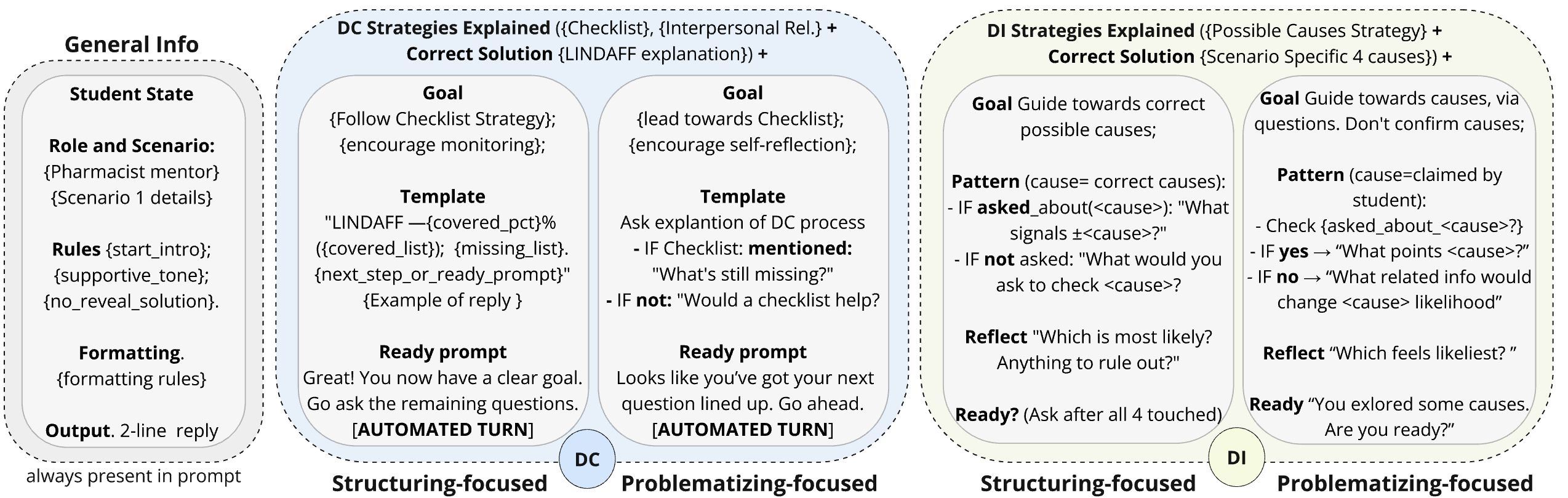}
  \caption{ Prompt Design for Data Collection and Data Interpretation per group}
  \Description{Prompts in Data Collection and Data Interpretation}
  \label{fig:prompts}
\end{figure*}

The overall architecture of the agents in the back-end of PharmaSim Switch is presented in Figure \ref{fig:arch-models}. From the LA point of view, it is grounded in constant monitoring of a student's learning trajectory by two rule-based models, while the LLM (implemented via the GPT-4o API with a temperature of 0.7 and default settings, using the system prompt and the most recent five interaction turns as context) allows natural language conversation with the pharmacist character.

\textbf{Client Inquiry and Research Module: Client Character.} The client character is implemented as a rule-based system using a predefined question–answer knowledge base that was expert-verified by teachers from a pharmacy technician vocational program in Switzerland. In addition, the rule-based \textit{Student Model Agent} tracks the data collection progress by monitoring the completion of the seven-item LINDAFF checklist during client–student interactions, which enables personalization of the Pharmacist Character. 

\textbf{Pedagogical Module: Pharmacist Character.} The agent behind the Pharmacist character is implemented as a state machine with two states: (1) data collection (DC), guiding students to gather a complete case by questioning the client, and (2) data interpretation (DI), supporting in analyzing the collected information and hypothesizing possible causes. The Student Model Agent manages state transitions: once all LINDAFF items are covered, it signals the Pharmacist Agent to shift from the data collection to the data interpretation state. To facilitate natural language, pedagogically grounded conversations with the pharmacist, a carefully crafted prompting strategy is tailored to each state and scaffolding method \cite{Reiser2004ScaffoldingWork} (Fig. \ref{fig:prompts}). More details about all of the prompts we used can be found in our repository \footnote{\url{https://github.com/epfl-ml4ed/PharmaSim-Switch/}}. In addition, the rule-based \textit{Student Model Agent} tracks the progress of data interpretation by monitoring mentioning of the four possible pre-defined causes during pharmacist-student interactions, enabling the agent to prevent the students from falling into the cognitive pitfall of premature closure \cite{Kumar2011}, allowing them only to progress to Diagnostic module if at least one possible cause was mentioned.

%%%%%%%%%% Method section %%%%%%%%%%%%%%%%%%%%%%%%%%%%%%%%%%%%%%%
\section{METHODOLOGY}

% Figure showing the overall design of the study
\begin{figure*}[t!]
 \centering
  \includegraphics[width=\linewidth]{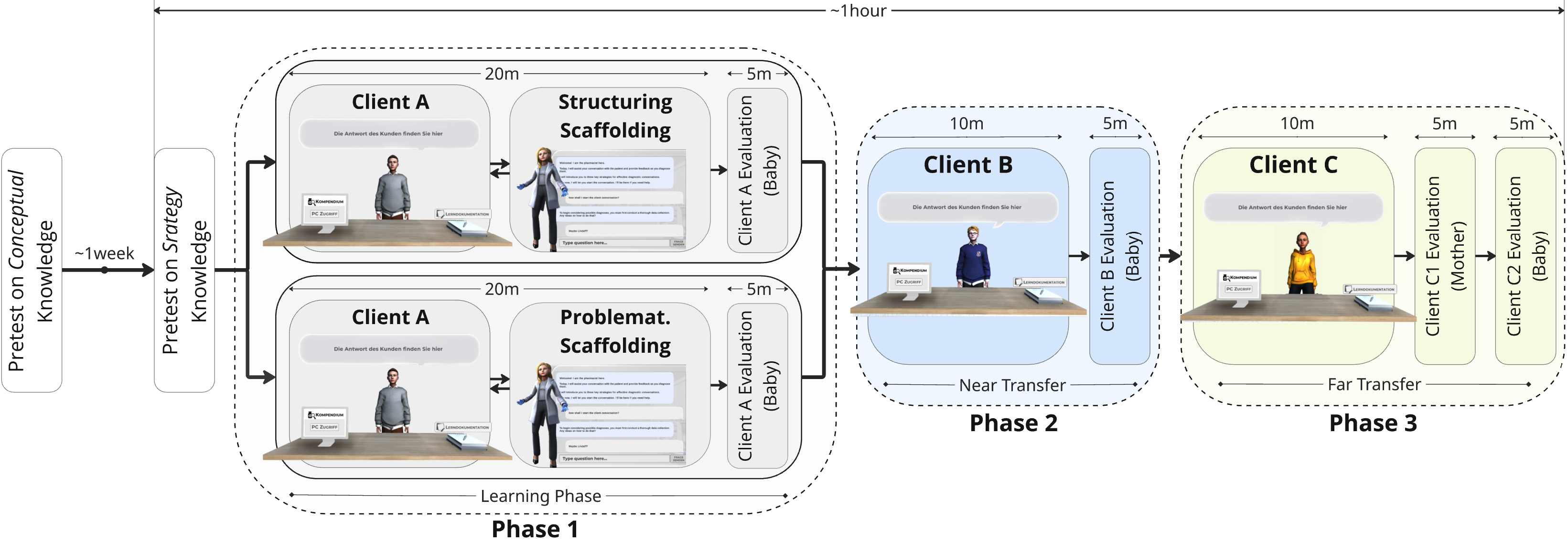}
  \caption{Experimental design. Pretests were followed by a learning phase with Structuring- or Problematizing-Heavy scaffolding (Phase~1) and two transfer phases (Phase~2–3).}
  \Description{Methodology}
  \label{fig:Methodology}
\end{figure*}

Our experimental design, as shown in Fig. \ref{fig:Methodology}, followed a between-group experimental structure, centered on students’ interactions with PharmaSim Switch, a scenario-based learning environment described in Section~\ref{sec:pharmasim_switch}. After completing pretests of theoretical and diagnostic strategy knowledge, students advanced through three phases that examined the impact of scaffolding approaches on learning outcomes and behavioral patterns. 

%%%%%% To be killed as a subsection. Scenario descriptions should be moved into procedures. %%%%%%%%%
\subsection{Study Procedure}
\label{sec:ExperimentalDesign}

Our study (see Fig.~\ref{fig:Methodology}) began with a pretest on conceptual knowledge and diagnostic strategies, followed by a learning session in which both groups participated in a diagnostic conversation. The subsequent phases then measured near-transfer and far-transfer again using simulated client scenarios from prior work \cite{radmehr2025pharmasimtext}, focusing on infant and maternal health issues that were ideally suited for examining both near and far transfer of diagnostic strategies.

\textbf{Pretests.} First, the conceptual knowledge pretest, developed by vocational school instructors for pharmacy technicians, was designed to assess baseline understanding of breastfeeding, infant nutrition, and common conditions in early infancy. Second, the strategy pretest required a brief 3–4 sentence response outlining an approach to understanding a pharmacy customer’s problem and providing effective advice, thereby measuring students’ prior knowledge of diagnostic strategies.

%The 21 short-answer items, provided by vocational pharmacy teachers, were scored on a 0–2 scale, and the results confirmed that there were no significant knowledge differences between groups before the study commenced.
%%% Description of study procedure

%\textbf{Strategy Knowledge Pretest.}
%The pretest aimed to ensure that the strategy knowledge levels of the experimental groups were similar. 
%Participants were asked to provide a brief, 3–4 sentence response outlining how they would approach understanding a pharmacy customer’s problem and offering effective advice.

\textbf{Phase 1 – Learning Session.}
In the first phase, students engaged in a diagnostic conversation (Client A) with a father concerned about his six-month-old baby experiencing diarrhea. Using diagnostic strategies, they were expected to evaluate four possible causes—teething, viral infection, dietary changes, and maternal medication—with dietary changes identified as the most likely explanation. During this session, participants had access to the Pedagogical Module.

\textbf{Phase 2 - Near Transfer.}
This phase assessed students’ ability to adapt their diagnostic reasoning to a slightly altered but related context (Client B), reflecting near-transfer. While the potential causes were identical to Scenario A, the conversation with the father indicated a different underlying cause—maternal antibiotic use—since the infant showed more severe viral symptoms, no recent dietary changes, and teething was unlikely.

\textbf{Phase 3 - Far Transfer.}
In the final phase, students engaged with a more complex case (Clients C1 and C2) that required them to apply their diagnostic reasoning in a substantially different context. The scenario involved a mother experiencing breastfeeding-related problems and concerns about her baby. Students were expected to identify three possible causes for the mother—breast engorgement, mammary gland infection, and cracked nipples—and three for the baby—bloating, diaper rash, and a mild cold. This phase represented far transfer, as it shifted the context from earlier cases by combining maternal and infant symptoms, thereby requiring students to extend and adapt their reasoning to a broader, more complex domain.

\textbf{Post-tests.}
Immediately after each client scenario (A, B, C1, C2), participants completed a short, scenario-specific post-test in the Diagnostic Module to assess their diagnostic reasoning. In this test, they were asked to identify possible causes, estimate their likelihood, and justify their reasoning. Importantly, during Phases 2 and 3, students did not have access to the Pedagogical Module, ensuring that the post-tests fairly measured both near and far knowledge transfer.

\subsection{Measurement and Analysis}
\label{subsec:measurement-analysis}
% Suggested structure
% Start with evaluation of diagnostic strategies as the first subsubsection (easier to describe). Put here what you currently have in 3.4.5 and then also directly merge 3.5.1.
% Have a second subsubsection: chatbot interactions, composed of three parts:
% PART-1: Surface-level features: describe the surface level features computed, why you selected them (references to literature) and what they mean
% PART-2: Scaffolding mechansim of chatbot: describe how you developed the scaffolding rubric (the rubric is a contribution, it needs to become clear that we developed it!). Describe the annotation and inter-rater agreement process.
% PART-3: Student interaction behavior: again, describe the dvelopment of the rubric (again, this is a contribution). When describing interrater agreement and annotation, you can refer to the process described for PART-2.

This section outlines how we defined and quantified all measures and our analytical approach. While students interacted with PharmaSim Switch, we logged all activity, including their questions to clients, the diagnostic hypotheses they generated and evaluated, and all interactions with the pharmacist. We then detail how pretests and post-tests were assessed, how diagnostic strategies were operationalized and coded, and how pharmacist–student interactions were annotated to evaluate alignment with the intended scaffolding approach.

\noindent\textbf{Pretest evaluations.}
The \textit{conceptual knowledge pretest} consisted of a 20-item multiple-choice questionnaire assessing theoretical knowledge of infant diarrhea and breastfeeding-related conditions (e.g., mastitis, mammary gland infection). The pretest showed acceptable internal consistency ($\alpha = 0.75$), and all items are provided in the project repository.
Each item was worth one point. The \textit{strategy knowledge pretest} assessing prior knowledge of diagnostic strategies was graded on a 0–4 scale based on four criteria: mentioning (i) the LINDAFF checklist, (ii) considering interpersonal relationships, (iii) generating a broad list of potential causes, and (iv) systematically prioritizing causes. Points were awarded for explicit and detailed references, partial credit for incomplete mentions, and zero for omissions.  

\noindent\textbf{Strategy performance evaluations.}
We assessed student performance on the \textit{checklist}, \textit{interpersonal relationships}, and \textit{data interpretation} strategies. The \textit{checklist strategy} was assessed based on adherence to the Checklist Strategy framework, where each of the seven categories corresponded to a required diagnostic question. A category was considered fulfilled if the relevant question appeared in the logged interactions. The Checklist Strategy score was calculated as the proportion of fulfilled categories, each contributing equally to the total score. The \textit{interpersonal relationships strategy} was assessed on a scale of 0 to 3 by checking whether students asked three relevant questions about another person in the scenario (mother in scenarios A and B, baby in scenario C). A category was fulfilled if at least one relevant question was asked, and the score was calculated as the proportion of fulfilled categories, quantifying consideration of interpersonal factors in diagnosis. The student's performance in the \textit{data interpretation strategy} was assessed by evaluating the number of potential causes identified and explanation clarity. Scores included correctly identifying all possible causes (0 to 4 in scenarios A and B, 0 to 3 for C1 and C2 separately) and correctly assessing the likelihood and rationale (scored on the same scale), while in the latter, partial explanations earned half points. To compare the scenarios, we normalize all the scores by calculating the percentage of the total.
%These three measures provide a multidimensional, structured evaluation of students' engagement, capturing the extent to which their actions aligned with diagnostic strategy relevance.

\noindent\textbf{Pharmacist-student discourse evaluation.}
To capture pharmacist and student behaviors, we consider three dimensions: surface-level indicators, a rubric characterizing the pharmacist agent's scaffolding moves, and a rubric capturing student behaviors and strategy use. Within each interaction, we defined a \emph{turn} as a contiguous run of utterances by the same party (student or pharmacist). Below, we describe the coding procedures per dimension and report inter-rater reliability.

%SURFACE
\textit{Surface-level indicators.} We analyzed surface-level features to capture interaction patterns between students and the pharmacist. Because students could alternate between the client and the pharmacist, we first examined switch behavior as the ratio of voluntary to total transitions in both directions (client$\rightarrow$pharmacist and pharmacist$\rightarrow$client). Each switch to the pharmacist, ending when the student returned to the client, defined one discussion. For each discussion, we measured the duration, number of switches, and average contributions of both the student and the pharmacist. An utterance was defined as a single sentence with at least one token, while a turn was a sequence of utterances by the same speaker without interruption. Based on these, we calculated ratios of student to pharmacist contributions, verbosity at both the utterance and turn levels, and interaction density as the number of utterances or turns per minute within the session’s phase span. Finally, we also evaluated surface features over the entire student–pharmacist interaction, including total utterances and turns by role, their ratios, and overall interaction density.

%CHATBOT RUBRIC
\textit{Pharmacist utterance evaluation rubric.} 
To evaluate the extent to which the pharmacist agent’s moves aligned with the intended pedagogical approach, we employed the framework of \citet{Reiser2004ScaffoldingWork}, which provides the theoretical foundation for distinguishing between structuring and problematizing scaffolding. In addition, each utterance was further categorized according to one of three targeted strategies: the checklist strategy, consideration of interpersonal relationships, and generation of possible causes. The categories with corresponding explanations and examples are presented in Table \ref{tab:pharmacist-rubric}. 

\begin{table*}[t!]
\scriptsize
    \centering
    \caption{Two-dimensional pharmacist utterance coding rubric combining framework of \citet{Reiser2004ScaffoldingWork} with underlying diagnostic strategies.}
    \begin{tabular}{p{0.17\linewidth}p{0.26\linewidth}p{0.20\linewidth}p{0.26\linewidth}}\hline
    \textbf{Strategies} &\textbf{Checklist }& \textbf{Interpersonal Relationships} & \textbf{Possible Causes}  \\ \hline
    \textbf{Structuring Scaffolding} \\ \hline
    
    Decomposing complex tasks (e.g., structuring the task solution into the plan) & 
    "Systematically recording symptoms is one of the most important steps for an effective diagnosis. " \newline
    %"The LINDAFF framework helps with this—it ensures that nothing important is overlooked. " \newline
    "Have you already used LINDAFF to collect data?"& 
    "We've focused on the baby so far — but is that the whole picture?"
    & 
    "Have you thought about possible causes that you could investigate further?" \newline
    "Now let's think about possible causes." \newline
    "What could be the cause of the baby's diarrhea?" \\ 
    
    Focusing effort (e.g., specific directions on what to do on a level) & "Great that you're considering localization! This helps to narrow down the problem even more precisely." \newline "Good thing you asked about the nature of the diarrhea! This helps to better understand the problem." & "Could the baby's condition also play a role? " \newline 
    "If something from the mother’s side were relevant, what would you pay attention to?" & "You asked about the intensity of the symptoms — good decision." \newline
    "You've brought up the baby's diet—good observation. "\\
    
    Monitoring progress (e.g., tracking progress, acknowledging what was already done) & "So far, you've covered about 66\% of the checklist—well done! Already asked: - Symptoms - Intensity - Location - Medical history Still open: - Duration - Allergies"& "We've focused on the baby so far..."& "You have narrowed down the possible causes. Now you are ready to make a probable diagnosis."\\ 
    
    \hline
    \textbf{Problematizing Scaffolding} \\ \hline
    
    Eliciting articulation (e.g., prompting for explanations) & "Do you think you've covered all the areas of the LINDAFF checklist, or is there still information you might be missing?" \newline "Would it help to use a checklist or framework to see what's still missing?" &  "Why do you think her condition might be relevant?" \newline "If something she takes affects the baby, how would we know?" & "What do you think of the father's answer? Could this help rule something out or confirm something?" \newline "You have considered different perspectives. Does one of them now feel more plausible than the others? Is there anything you would now question or rule out?"\\
    
    Eliciting decision (e.g., prompting for self-decision) & "Want to complete the rest of the checklist?" \newline
    %"Do you think you are done with LINDAFF?" \newline
    "Do you feel that you have covered all relevant areas of the LINDAFF scheme, or are there still open questions you would like to clarify before considering the root causes?" & "Are we sure we’ve collected information from all relevant people involved in the baby’s care?" 
    &“Which cause seems most likely to you so far?“ \newline
    “Could several factors play a role at the same time? Are there any explanations that you would now rule out?”
    %“Would you say teething is a likely or unlikely cause?”
    \\
    
    Pointing at gaps and disagreements (e.g., showing contradiction in logic) & "You've already asked about fever." & "You have thought of mothers medication intake, but never asked anything about the mother..." & "You rated viral ‘unlikely’ due to no fever, then called it ‘very likely’"\\\hline
    
    \textbf{Affirmative responses (e.g., neutral acknowledgment, positive encouragement)} & \multicolumn{3}{p{0.75\linewidth}}{  
    "Great that you're considering the location! That helps narrow down the problem even more precisely." \newline 
    "Hmm.. interesting that you think that way"} \\
    
    \textbf{Pharmacist Mistakes} & \multicolumn{3}{p{0.75\linewidth}}{Errors in Monitoring; Encouraging Medical Advice; Replying to questions meant for the client;}\\ 
    \bottomrule
    \end{tabular}
    \label{tab:pharmacist-rubric}
\end{table*}

\textit{Student utterance evaluation rubric.} 
To analyze levels of students' engagement in interactions with the pharmacist, we employed the \textit{Active–Constructive–Interactive} (ICAP) framework \cite{Chi2009}. This framework distinguishes between different modes of engagement: Active (engaging through simple actions such as note-taking), Constructive (creating new ideas or connections beyond the given information), and Interactive (co-constructing knowledge through reciprocal dialogue). Each response was further classified as \textbf{Correct} or \textbf{Incorrect}, capturing whether the contribution aligned with the target diagnostic reasoning strategies and the collected data.
%with a dimension of \textit{correctness}. Along the ICAP dimension, student turns were coded as \textbf{Active} (short answers without elaboration), \textbf{Constructive} (generation of new ideas not explicitly provided by the chatbot), or \textbf{Interactive} (building on the chatbot’s ideas through reasoning or co-construction). 
The resulting two-dimensional coding scheme with category clarifications and examples is presented (Table~\ref{tab:aci-correctness}). %enabled us to distinguish between superficial replies, novel but misguided contributions, and collaborative reasoning episodes that advanced the diagnostic process. 

\begin{table*}[t!]
\scriptsize
    \centering
    \caption{Two-dimensional student labeling rubric combining the Active–Constructive–Interactive (ICAP) framework with correctness.}
    \begin{tabular}{p{0.07\linewidth}p{0.43\linewidth}p{0.43\linewidth}}\hline
    &\textbf{Correct }& \textbf{Incorrect}  \\ \hline
    \textbf{Active} & Short correct answer without any elaboration (e.g., "yes"; "no") \newline 
    Repetition of factual knowledge given in the simulation (e.g., "The baby is five months old.")
    & Short incorrect answer without any elaboration (e.g., "Might be"; "I believe so") \newline
    Attempt to reveal the correct solution (e.g., ``What are the possible causes?'') \newline
    Non-progressing reply (e.g., ``leave me alone...'') \newline
    Expressing confusion (e.g., ``I don't know'')
    \\       
    \textbf{Constructive} & Introducing correct new ideas (e.g., a reasonable possible cause), not related to the previous conversation (e.g., “A viral infection could explain the sudden diarrhea.”, “Could it be that the baby doesn’t tolerate puree?”) &  Introducing non-productive incorrect new ideas (e.g., "I will send him to the doctors", Maybe the baby has intolerance to breast milk)\\
    \textbf{Interactive} & Introducing correct new ideas (e.g., a reasonable possible cause), reasoning with the help of scaffolding, built on ideas of the pharmacist (e.g., “Yes, it could be due to the mother’s medication as this also affects breast milk”)  &
    Introducing incorrect new ideas, built on ideas of the pharmacist \newline
   (e.g., pharmacist: ``could it be diet change?'', student: ``yes, I think it is due to bimbosan'')\\
    \bottomrule
    \end{tabular}
    \label{tab:aci-correctness}
\end{table*}

The annotation was conducted as follows. First, the pharmacist-student interactions were automatically segmented into turns and utterances (sentence-based). Two raters (paper authors) with expertise in learning sciences conducted the annotation using the above schemes. As a pilot, two complete pharmacist-student interactions were coded and discussed to refine the coding scheme. During these conversations, the clarifying example utterances were added to both rubrics. Following this calibration, ten additional interactions (five per scaffolding approach, totaling 328 pharmacist and 90 student utterances) were independently annotated. %Together, the pilot and reliability sets covered approximately 16\% of the dataset. 
For the pharmacist’s utterances, inter-rater agreement was assessed across both scaffolding categories and their corresponding strategy labels, yielding an acceptable level of reliability (Cohen’s k = 0.66). For students’ utterances, agreement was calculated across ICAP categories and correctness dimensions, and it also reached an acceptable level (Cohen’s k = 0.62). After all the disagreements were discussed and resolved, the coding of the remaining data was done by a single rater.

%Interrater agreement was assessed on a subset of ten students ($n = 90$ coded utterances). The two raters reached 70.0\% agreement. Cohen’s $\kappa$ indicated substantial agreement  between the raters ($\kappa = 0.62$).  
%The two raters reached 69.8\% agreement. Cohen’s $\kappa$ indicated substantial agreement between the raters (\(\kappa = 0.66\)), suggesting that the applied coding scheme was reliable. Importantly, agreement was calculated at the justification label level, which required raters to match not only the overarching scaffolding category but also the specific subcategory and diagnostic strategy. This provided a more precise assessment of labeling consistency. Disagreements were resolved through discussion.

\subsection{Analytic approach}
For \textbf{RQ1}, all pharmacist utterances were aggregated at the student level, and the proportion of turns in each scaffolding mechanism (structuring, problematizing, affirmative responses, and mistakes) and each diagnostic strategy (checklist, interpersonal, and possible causes) was calculated relative to the total pharmacist utterances in that student’s session. These produced per-student percentage scores served as the unit of analysis. We compared the groups using independent-samples $T$-tests with multiple comparisons across scaffolding categories and diagnostic strategies, adjusting for the Bonferroni method. 

For \textbf{RQ2}, we analyzed how the interaction of clients (A, B, C1, and C2) and instruction group (Structuring-heavy vs.\ Problematizing-heavy scaffolding) influenced students' diagnostic strategy scores. We used a Mixed Linear Model (MLM)\footnote{\textit{lme4} and \textit{afex} R packages \label{fn:r}}  appropriate for a between-group experimental design with repeated measures.  Students' two pretest results were included as fixed effects, while individual variation between students was modeled as a random intercept. We fit separate MLMs for each diagnostic strategy. For the \textit{checklist} and \textit{interpersonal relationship} strategies, scores were calculated per client (A, B, and C). For the \textit{post-test} (data interpretation), scores were calculated per post-test (A, B, C1, and C2). In each model, we conducted follow-up pairwise comparisons of estimated marginal means using the \textit{emmeans} package, with Tukey correction for multiple comparisons. 

%\subsubsection{RQ3 evaluation}
For RQ3, all student utterances were aggregated at the student level, and the proportion of turns in each ICAP  (Active, Constructive, and Interactive) and correctness dimension was calculated relative to the total student utterances in that student’s session. These produced per-student percentage scores served as the unit of analysis. We compared the groups using independent-samples $T$-tests with multiple comparisons adjusted using the Bonferroni method. 

% Participants
\subsection{Participants}
Pharmacy technician apprentices (N = 63, 60 female) participated, reflecting the profession's gender distribution in Switzerland. All were second-year apprentices in a three-year vocational program with prior coursework in pharmaceutical sciences. Students (n = 6) who were absent or failed to complete the pretest were excluded from analysis. Participation was voluntary and integrated into routine training, with informed consent obtained (for minors, from their parents). Participants were randomly assigned to one of two instructional conditions: Structuring-heavy Scaffolding (n = 31) or Problematizing-heavy Scaffolding (n = 32). Post-hoc sensitivity analyses indicated that this sample size provided adequate power ($1-\beta > .80$) to detect medium effect sizes ($d \approx 0.65$) in independent-samples $T$-tests and medium interaction effects ($f \approx 0.25$) in repeated-measures designs, which aligns with prior scaffolding and learning analytics research. The study was approved by the university's ethics committee (HREC 636-2024).

%%%%%%%%%Result Section%%%%%%%%%%%%%%%%%%%%%%%
\section{RESULTS}
%short intro
%\textbf{Verification of randomization}.
We verified randomization by checking the pretest performance of the participants. For the domain knowledge pretest, scores met normality assumptions (Shapiro–Wilk, p > .05), and no group differences were found ($t(63) = -0.41, p = .682$). For the strategy knowledge pretest, scores violated normality (Shapiro–Wilk, p < .001); a Mann–Whitney $U$ test showed no significant group differences ($U = 498.5, p > .05$). Below, we present the study findings, organized according to the three research questions.
%Both groups primarily demonstrated familiarity with the checklist strategy, with no other methods being reported.
\vspace{-5pt}
\subsection{RQ1: Pharmacist Agent Alignment with the Intended Pedagogical Approach}
%\subsubsection{Scaffolding Mechanisms, Strategy Distributions, and Their Interplay}
%\textbf{First Analysis: Scaffolding Mechanisms.}
% Figure displaying the scaffolding distribution
\begin{figure*}[t!]
    \includegraphics[width=1\textwidth]{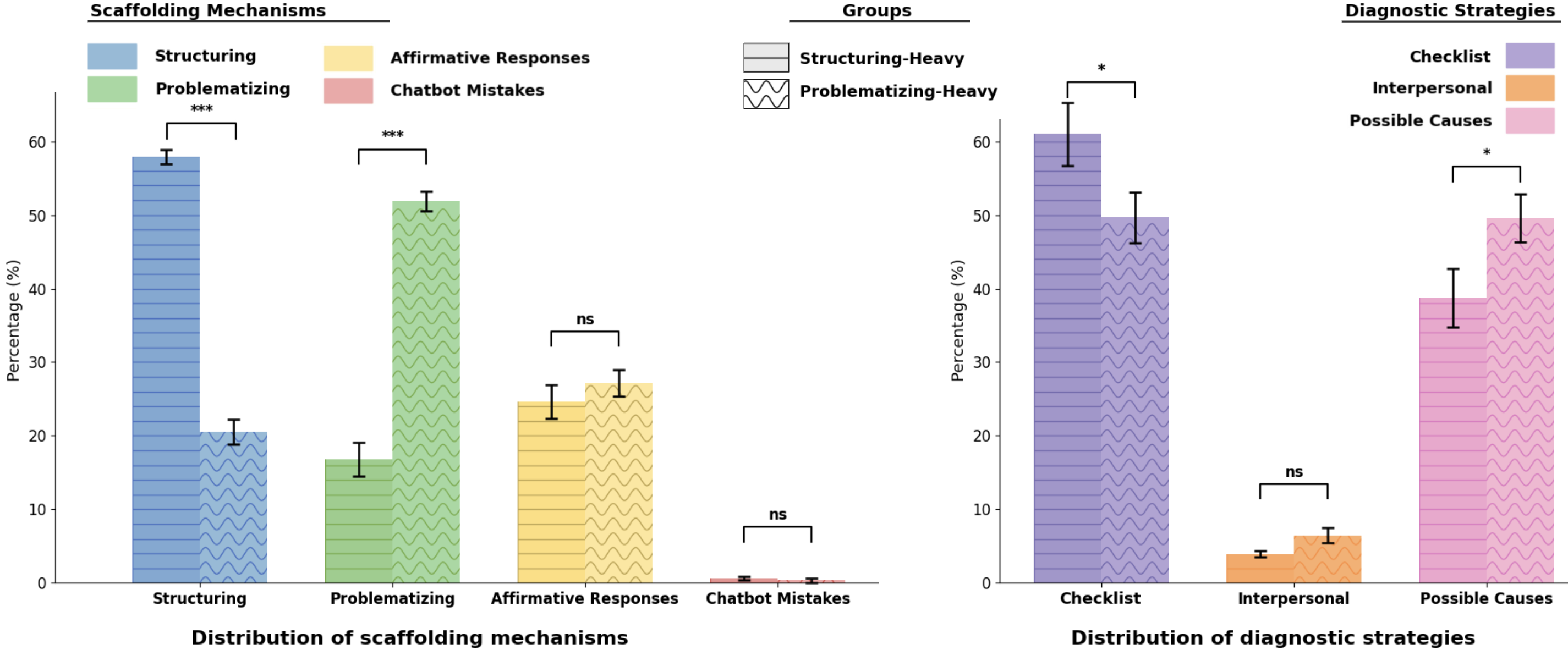}
    \caption{RQ1: Distribution of scaffolding mechanisms by group (Left). 
    Distribution of diagnostic strategies by group (Right). Mean percentages with standard errors and pairwise tests }
    \label{fig:PercentagesStructuringProblematizing}
\end{figure*}

% Sankey-plot showing flows from scaffolding to sub-scaffolding to diagnostic strategies
\begin{figure*}[t]
    \centering
    \includegraphics[width=0.9\linewidth]{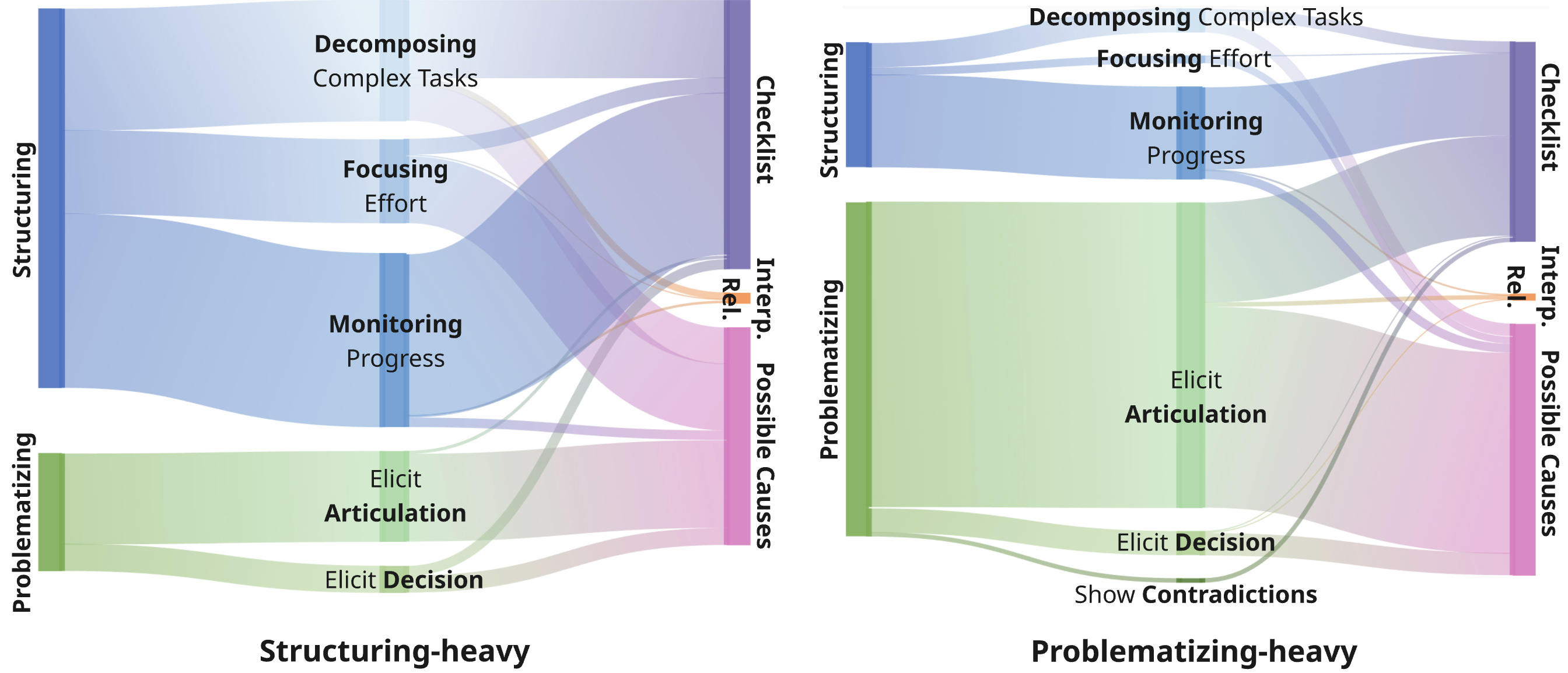}
    \caption{RQ1: Flows from scaffolding category to subcategory and diagnostic strategies per group.}
    \label{fig:SankeyScaffoldingStrategies}
\end{figure*}

First, we compared the distribution of scaffolding categories between the two conditions. The results demonstrate a clear divergence in scaffolding practices (Fig. \ref {fig:PercentagesStructuringProblematizing}, left). T-tests showed that structuring occurred significantly more often in the Structuring-heavy group and problematizing in the Problematizing-heavy group (both p < .001), with no differences for affirmative responses or pharmacist mistakes (both p > .05). Across both conditions, pharmacist mistakes produced by the LLM were rare (approximately 1--2\% of mentor utterances) and strictly pedagogical in nature (e.g., prematurely revealing answers or responding to prompts intended for the client rather than the mentor). While such mistakes may reduce learning value, they do not pose risks to real patients in this simulated setting.

Second, we analyzed how the pharmacist agent conveyed diagnostic strategies (Figure~\ref{fig:PercentagesStructuringProblematizing}, right). The T-tests confirmed that there was no significant difference between the two groups for all three strategies (all p > .05). Both agents emphasized the Checklist and Possible Causes strategies. In contrast, the Interpersonal Relationship strategy was not frequent. These results confirm that the prompting strategies effectively guided the pharmacist agent toward the intended scaffolding approach.

%Figure on scores per scenario and client
\begin{figure*}[t!]
    \centering
    \includegraphics[width=\linewidth]{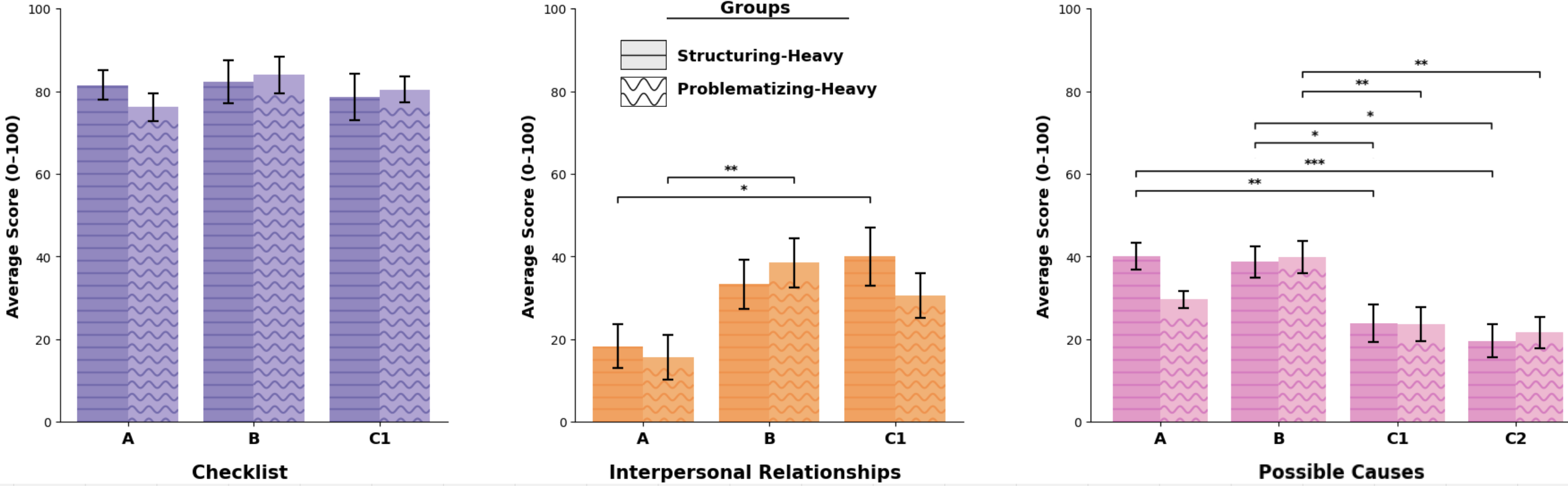}
    \caption{RQ2: Distribution of diagnostic strategy evaluation scores per scenario/client. }
    \label{fig:RQ1}
\end{figure*}

\begin{table*}[t!]
\scriptsize
    \centering
    \caption{The qualitative examples per sub-scaffolding category..}
    \begin{tabular}{p{0.15\linewidth}p{0.39\linewidth}p{0.39\linewidth}}
     \hline
          \textbf{Scaffolding Subcategory} &\textbf{Structuring-heavy implementation }& \textbf{Problematizing-heavy implementation}  \\ \hline
          Structuring Scaffolding\\\hline
          Decomposing& “Let’s split this into symptoms, duration, and severity to cover the checklist efficiently.”& “What smaller questions would help you untangle this case?”\\
          Focusing Effort & “Focus next on clarifying intensity to advance the checklist.” & “Which single question would most reduce uncertainty now?”\\
          Monitoring& %“You’ve completed about 50\%—done: symptoms, duration; pending: intensity, history.” 
          “I see that you are already applying Checklist Strategy principles.  So far, you've covered about 33\% of the checklist!  Already asked: Symptoms, Intensity. Still open: Duration, Allergies, Location, Medical history.” & %“You’re part-way through; what key piece is still missing?”
          “You have already covered many aspects of the Checklist Strategy checklist, such as symptoms, intensity, and duration. Is there anything else you would like to investigate further to better narrow down possible causes of the diarrhea?”\\
    \hline
    Problematizing Scaffolding\\\hline
    Elicit Articulation & “How likely is the new diet the cause at this point?”
& “Explain why the new diet might or might not account for these symptoms.” \\
Elicit Decisions & “Given the evidence so far, which cause is most likely?”
& “Which hypothesis do you commit to now, and why?”\\
 Show Contradictions& - &“Yes, Localization is an important question to ask, but you have already covered it.”\\
\hline
    \end{tabular}
    \label{tab:QualitativeComparison}
\end{table*}

Next, we analyzed how the higher-level scaffolding pedagogy (structuring vs.\ problematizing) manifested in their subcategories and how these were associated with the three diagnostic strategies Fig. \ref{fig:SankeyScaffoldingStrategies}). In both groups, structuring scaffolding was predominantly expressed through Monitoring ($\mu^{Struc} = 45.8\%$, $\mu^{Probl} = 74.3\%$), followed by Focusing effort ($\mu^{Struc} = 32\%$, $\mu^{Probl} = 19.3\%$) and Decomposing complex tasks ($\mu^{Struc} = 22\%$, $\mu^{Probl} = 6.4\%$). Similarly, problematizing scaffolding was predominantly expressed through Elicitation of Articulation ($\mu^{Struc} = 77\%$, $\mu^{Probl} = 91.4\%$), followed by Elicitation of Decision ($\mu^{Struc} = 23\%$, $\mu^{Probl} = 27.2\%$). The Problematizing-heavy agent rarely used the Show Contradiction ($\mu^{Probl} = 1.3\%$), while the Structuring-heavy agent never used it. Next, we examined, for each diagnostic strategy, how frequently the pharmacist agent employed the two scaffolding approaches. 

In the Structuring-heavy group, to support the Checklist strategy ($\mu^{Struc} = 95\%$) and the Interpersonal relations strategy ($\mu^{Struc} = 100\%$), the pharmacist used a structuring approach almost exclusively. By contrast, when supporting the Possible causes strategy, structuring ($\mu^{Struc} = 51.8\%$) and problematizing were applied with nearly equal frequency. In the Problematizing-heavy group, the possible causes and interpersonal relationship strategies were supported primarily through problematizing scaffolding ($\mu^{Probl} = 88.3\%$ and $\mu^{Probl} = 75\%$, respectively), while when supporting mastering the checklist strategy, the Problematizing-heavy pharmacist used both structured and problematizing approaches with a near equal rate ($\mu^{Probl} = 47.3\%$). These findings suggest that while procedural strategies, such as the checklist and interpersonal relations, are best supported through structuring, more generative strategies, like the Possible causes, benefit from a balanced combination of structuring and problematizing scaffolding.

Finally, we qualitatively examined how the pharmacist's scaffolding behaviors unfolded in practice. Interestingly, although the same categories of labels could occur in both conditions (see Table~\ref{tab:QualitativeComparison} for representative examples), they were realized differently in the Structuring-heavy versus Problematizing-heavy conditions. For instance, for the Monitoring subcategory, the Structuring-heavy agent emphasized coverage and completion, whereas the Problematizing-heavy agent framed monitoring in a more exploratory way.

% R2: Effect of chatbot instruction on student learning

\subsection{RQ2: Effect of pharmacist instruction on student learning}
To test the effects of an interaction with the Structuring or Problematizing pharmacists on student learning, we run an MLM model for each diagnostic strategy. The results of the MLM are presented in our repository. The analysis of the MLM models detected no effect of either strategy or conceptual knowledge pretests, the experimental group condition, or the interactions between scenarios and experimental groups across all strategies. However, there was a significant effect of the diagnostic scenarios for the interpersonal relationship (p < .05) and the possible causes strategies (p < .05). The results of post-hoc pairwise comparisons are presented in Fig.~\ref{fig:RQ1}.  

% Heatmap plots linking interaction behavior and correctness
\begin{figure*}[t!]
    \centering
    \includegraphics[width=\linewidth]{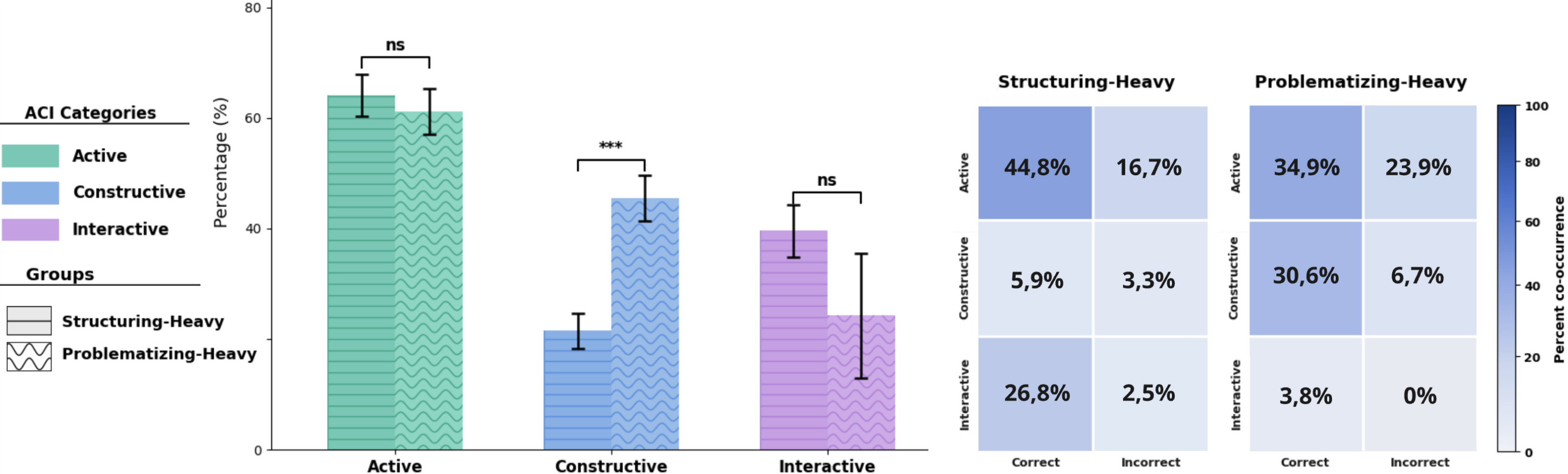}
    \caption{RQ3: ICAP behavior by group (Left). Heatmaps linking ICAP categories to correctness (Right)}
    \label{fig:heatmap}
\end{figure*}

The results revealed no significant differences between the Structuring and Problematizing groups for all clients (p > .05) and across all scenarios. Regarding the pairwise comparisons between the clients, there was a significant difference in interpersonal relation strategy scores for Problematizing Group between client A ($\mu_{A}^{Probl}$ = 15.6, $SE_{A}^{Probl}$ = 5.39) and  Client B ($\mu_{B}^{Probl}$ = 38.5, $SE_{B}^{Probl}$ = 6.01), p < .01. And for the Structuring Group between Client A ($\mu_{A}^{Struct}$ = 18.3, $SE_{A}^{Struct}$ = 5.32) and Client C1 ($\mu_{C1}^{Struct}$ = 40.0, $SE_{C1}^{Struct}$ = 7.06), p < .05 (see Fig.~\ref{fig:RQ1}). These findings suggest that the students in both groups had successfully mastered the interpersonal relationship strategy, with the Structuring Group being able to apply it even in the far-transfer scenario. Furthermore, students in both groups demonstrated a significant degradation in scores for the possible causes strategy in Scenarios B ($\mu_{B}^{Probl}$ = 39.8, $SE_{B}^{Probl} = 3.91$, $\mu_{B}^{Struct}$ = 38.7, $SE_{B}^{Struct} = 3.84$) with both C1 ($\mu_{C1}^{Probl}$ = 23.7, $SE_{C1}^{Probl} = 4.08$, $\mu_{C1}^{Struct}$ = 23.9, $SE_{C1}^{Struct} = 4.55$) and C2 ($\mu_{C2}^{Probl}$ = 21.6, $SE_{C2}^{Probl} = 3.85$, $\mu_{C2}^{Struct}$ = 19.6, $SE_{C2}^{Struct} = 4.01$), all p < .01 (besides p < .05 for the difference between Clients B and C1 for the Structuring Group). In addition, for the Structuring Group only, there was a significant degradation in possible causes strategy scores between Client A ($\mu_{A}^{Struct}$ = 29.7, $SE_{A}^{Struct} = 2.06$) and both C1 ($\mu_{C1}^{Struct}$ = 23.9, $SE_{C1}^{Struct} = 4.55$), p < .01 and C2 ($\mu_{C2}^{Struct}$ = 19.6, $SE_{C2}^{Struct} = 4.01$), p < .001 (Fig.~\ref{fig:RQ1}). These findings suggest that both groups had difficulty applying the possible causes strategy to a more challenging scenario (Clients C1 and C2). There were no significant differences between the Clients in Checklist strategy scores for both groups (Fig.~\ref{fig:RQ1}).

% Give a short introductory sentence: To test the effects of a structuring-heavy or problematizing-heavy chatbot on student learning, we...
% Then introduce the results of the general MLM (not testing for interaction effects between pre-test and condition), you can reference the table. Both pretests need to be included as covariates into the model
% Then introduce ther results of the post-hoc comparison (Figure 8). 
% After that discuss the findings in a structured way.

%1) MLM almost exactly the same as in the AIED paper (only difference is the inclusion of the conceptual pretest and the strategy pretest as a covariate)
%2) Deep-dive analysis questioning whether there is an interaction between prior conceptual knowledge and chatbot type or prior strategy knowledge and chatbot type.

% Table reporting the MLM results

% RQ3: How does student learning differ between different types of scaffolding?
\subsection{RQ3: Student Interaction Behavior}

% First result - surface level interaction features
%\subsubsection{Surface-Level Features}
%\label{sec:results-surface}

%Following the analysis order presented in the methodology, we report the main effects and refer readers to Table~\ref{tab:surface_all_compact_sd} for full statistics. 

We first analyzed surface-level features of the interaction. Students in the Structuring-heavy condition switched to the pharmacist more often than in the Problematizing-heavy condition (6.19 vs.\ 4.82 per session, $p = .039$), with a marginally higher client$\to$pharmacist ratio (0.629 vs.\ 0.482, $p = .054$).  At the \emph{discussion} level, Structuring-heavy interactions tended to be shorter (54.06\,s vs.\ 74.48\,s, $p = .057$), whereas Problematizing-heavy interactions elicited more student activity, with more utterances (2.39 vs.\ 1.50, $p = .009$), words (4.78 vs.\ 2.99, $p = .009$), and turns (3.03 vs.\ 2.05, $p = .028$). Pharmacists also contributed more turns in these sessions (2.94 vs.\ 2.00, $p = .008$).  In terms of \emph{verbosity}, Structuring-heavy sessions produced longer turns (32.80 vs.\ 27.33 words, $p < .001$) but shorter utterances (11.17 vs.\ 15.50 words, $p < .001$). Participation ratios were lower in Structuring-heavy sessions, both per discussion (0.21 vs.\ 0.37, $p < .001$) and across sessions (utterances: 0.21 vs.\ 0.37, $p < .001$; turns: 0.66 vs.\ 0.74, $p = .029$).  Finally, \emph{interaction density} was higher in Structuring-heavy sessions, within pharmacist spans (22.59 vs.\ 14.24 utterances/min, $p = .002$) and across full sessions (2.51 vs.\ 1.86 utterances/min, $p = .002$). Over entire sessions, pharmacists also produced more utterances in Structuring-heavy (35.13 vs.\ 22.06, $p < .001$), whereas student totals did not differ significantly (table available in our repository).

% Second: discussion of interaction behavior
%\subsubsection{Student Interaction Labeling}
%\subsubsection{Passive}
%Passive behaviour was more frequent in the Structuring-heavy group (44.7\%) compared to the Problematizing-heavy group (36.7\%). This difference was statistically significant, with Structuring-heavy scaffolding eliciting more passive choices ($p = .041$).
% Figure showing ICAPpercentages for RQ3
%\begin{figure}[t]
%    \hspace{0cm}
%    \includegraphics[width=0.8\textwidth]{Images/ICAP.png}
%    \caption{RQ3: Distribution of ICAP (Active–Constructive–Interactive) responses by group, mean percentages with standard errors (Left). Mean percentages with standard errors (Right) (ns = not significant) (*** $p<.001$; ns = not significant). }
%    \label{fig:ACI}
%\end{figure}

Next, we compared students’ interaction behaviors across scaffolding conditions using the Active-Constructive-Interactive (ICAP) framework (Figure~\ref{fig:heatmap}). In the Structuring-heavy group, responses were mostly Active (64.0\%), with fewer Constructive (21.5\%) and Interactive (39.6\%) contributions. By contrast, the Problematizing-heavy group showed a larger share of Constructive responses (45.5\%) and fewer Interactive ones (24.3\%), while Active responses were similar (61.2\%). Mean percentages with standard errors confirmed these patterns: Constructive responses were significantly more frequent in the Problematizing-heavy group ($p < .001$), whereas no significant differences emerged for Active ($p = .329$) or Interactive responses ($p = .215$). Overall, both conditions promoted Active engagement, but Problematizing-heavy scaffolding fostered more constructive reasoning.  Finally, we assessed the correctness of student responses. Both groups showed similar proportions of correct answers (Structuring: 77.4\%, Problematizing: 74.8\%), with no significant differences for either correct ($p = .482$) or incorrect responses ($p = .317$). Thus, the scaffolding condition affected the type of engagement but not response accuracy.

% Second analysis: were student answers in the chatbot correct?
%\subsubsection{Correct Incorrect}
%\begin{figure}[t!]
%    \hspace{0cm}
%    \includegraphics[width=1\textwidth]{Images/correct.png}
%    \caption{RQ3: Distribution of correctness (Correct vs. Incorrect) by group. Stacked proportions (Left). .}
%    \label{fig:correct}
%\end{figure}

% Third analysis: 
%\subsubsection{Linking ICAPto Correctness}
%In a third analysis...Figure~\ref{fig:sankey2} illustrates how student responses in the Active--Constructive--Interactive (ACI) framework translated into correct and incorrect answers. In the Problematizing-heavy group, constructive responses were frequently associated with correct answers, while interactive responses were rare and mostly linked to incorrect answers. Active responses were the dominant category and split between correct and incorrect outcomes. In the Structuring-heavy group, active responses again dominated and showed a mix of correct and incorrect outcomes. Interactive responses were more prominent than in the Problematizing-heavy group and often flowed into correct answers, although some were also linked to incorrect ones. Constructive responses occurred less often overall but leaned toward correctness when present. Taken together, the Sankey diagrams indicate that Problematizing-heavy scaffolding encouraged constructive reasoning aligned with correctness, whereas Structuring-heavy scaffolding elicited more interactive contributions, which were more numerous and sometimes correct but not consistently so.

The patterns of student engagement differed noticeably between the two scaffolding conditions (Fig.~\ref{fig:heatmap}). In the Structuring-heavy group, most student utterances were correct, with nearly half classified as Active (44.8\%), more than a quarter as Interactive (26.8\%), and a smaller share as Constructive (5.9\%). Incorrect contributions were comparatively infrequent, comprising Active (16.7\%), Constructive (3.3\%), and Interactive (2.5\%) responses. In contrast, in the Problematizing-heavy group, students engaged more through constructive reasoning, as reflected in a higher share of Constructive Correct responses (30.6\%). This increase, however, coincided with declines in Active Correct (34.9\%) and Interactive Correct (3.9\%). At the same time, accuracy dropped overall, with Active Incorrect responses rising to 23.9\% and Constructive Incorrect to 6.7\%. Overall, structuring scaffolding supported more accurate active and interactive participation, whereas problematizing scaffolding promoted deeper constructive engagement but also led to more frequent errors.

\section{DISCUSSION AND CONCLUSIONS}
% Short intro sentence
In this paper, we investigated how LA- and LLM-powered pharmacist agents can support the different scaffolding approaches of structuring versus problematizing, and how these differences shape learners' interaction behavior in a scenario-based simulation environment.

%Results for RQ1
Regarding \textbf{RQ1}, we found that LA- and LLM-powered agents effectively instantiated theoretically grounded scaffolding distinctions. The Structuring-heavy pharmacist emphasized task decomposition, focused attention, and checklist monitoring, while the Problematizing-heavy pharmacist elicited articulation, decisions, and contradictions, consistent with prior work \cite{Reiser2004ScaffoldingWork}. Notably, structuring placed stronger emphasis on the checklist strategy, whereas problematizing supported reasoning about the possible causes. This alignment between the scaffolding approach and strategy emphasis was emergent rather than explicitly designed. 
%Results for RQ2
In \textbf{RQ2}, both scaffolding conditions supported the development of diagnostic reasoning strategies, with no significant differences in overall performance or knowledge transfer. This may reflect domain complexity and learner expertise, as vocational students could benefit from both scaffolds in different ways, while our transfer tasks may not have been sensitive enough to distinguish procedural from conceptual gains \cite{barnett2002when,Graber2012}. We observed scenario-specific effects, particularly when applying the interpersonal relationships strategy in far-transfer cases, suggesting that both versions of PharmaSim Switch helped learners generalize aspects of diagnostic reasoning. 
%Results for RQ3
For \textbf{RQ3}, we analyzed students' interaction behaviors to examine how scaffolding influenced engagement. The Structuring-heavy version produced shorter but more frequent interactions, with higher density and more pharmacist-led contributions. In contrast, the Problematizing-heavy version elicited longer, student-driven exchanges and significantly more Constructive responses \cite{Chi2009}, consistent with work linking problematizing scaffolds to productive struggle and deeper engagement \cite{Quintana2004AInquiry}, though sometimes less accurate. Structuring was associated with more correct responses, but primarily with Active or Interactive engagement, rather than Constructive elaboration. These results suggest that scaffolding types foster distinct learner behaviors, which may be more or less desirable depending on the instructional goals. %From a learning analytics perspective, fine-grained interaction data from SBL simulations can reveal such process-level differences and inform the design of AI-supported scaffolding.

% Implications
Our findings have several \textbf{implications} for LA researchers and educational stakeholders. For educators and instructional designers, our study provides evidence-based guidance on operationalizing different scaffolding approaches using LA- and LLM-based agents to support diagnostic reasoning. 
%By analyzing student–agent interactions through established pedagogical frameworks \cite{Reiser2004ScaffoldingWork, Chi2014ICAP}, we demonstrate how scaffold design influences both correctness and the type of student engagement, providing insights for shaping real-time instructional interventions. Our work extends beyond merely outcome-based evaluation by demonstrating how discourse analysis and exploratory analytics can be utilized to examine the processes and patterns of LA- and LLM-agent and learner behaviors using fine-grained, theory-informed interaction data collected from our simulation. We further show how LA can be leveraged to evaluate the pedagogical alignment of AI-based tools in real-world learning environments, which is crucial in human-centered learning analytics \cite{alfredo2024human, yan2024generative}. Our findings, therefore, suggest a promising direction for combining LA and generative AI to provide evidence-based, pedagogically aligned personalized support~\cite{Khosravi_Shibani_Jovanovic_A_Pardos_Yan_2025}. 
Moreover, our findings raise new questions for LA researchers about how to balance productive struggle with accuracy when designing adaptive support systems, suggesting future work on dynamic scaffolding that adjusts in real time based on behavioral indicators collected from learners. Several \textbf{limitations} align with common challenges in LA research \cite{10.1145/3636555.3636884}. First, the study examined pharmacy apprentices in a single national context, limiting generalizability; replication across domains and learner populations is needed. Second, outcomes were measured only immediately, without evidence on retention or long-term transfer; future work should include delayed post-tests. Third, although the LLM was prompted to provide theory-driven scaffolding, generative variability produced uneven strategy coverage (e.g., limited support for interpersonal reasoning), which may constrain learning. Future research should balance strategy emphasis while preserving adaptive variation and monitor how LA-/LLM-driven agents may implicitly reinforce particular instructional behaviors.
% Limitations and future work
%Several \textbf{limitations} should be noted, consistent with typical challenges in LA research \cite{10.1145/3636555.3636884}. First, the study involved pharmacy apprentices in a single national context, which limits generalizability; therefore, replication across different domains and learner populations is needed. Second, outcomes were measured immediately, without follow-up on retention or long-term transfer, which future work should address with delayed post-tests. Third, while LLMs were prompted to enact theory-driven scaffolding, the generative variability led to uneven coverage. for example, limited guidance on interpersonal reasoning, which potentially constrains learning. Future research should explore ways to balance strategy emphasis while preserving adaptive variation, and more broadly, ensure careful monitoring of how LA- and LLM-driven agents may implicitly reinforce certain instructional behaviors.

\noindent \textbf{Acknowledgments.}
This study was funded by the Swiss State Secretariat for Education, Research and Innovation SERI.

%In summary, our study shows how learning analytics can be used to evaluate and compare the pedagogical roles of LLM-driven agents in scenario-based learning. By linking scaffold type to learner behaviors and reasoning strategies, we highlight the importance of aligning AI support with instructional goals, contributing both theoretical and practical insights to the learning analytics literature for the design and evaluation of adaptive learning technologies in real-world scenrio-based learning settings.

\renewcommand{\refname}{REFERENCES}
\bibliographystyle{ACM-Reference-Format}
\bibliography{references}

\end{document}